\documentclass[12pt]{iopart}

\usepackage{iopams}
\usepackage{setstack}
\usepackage{amscd}
\usepackage{graphicx}
\usepackage{hyperref}
\usepackage{bm}
\usepackage{psfrag}

\newcommand{\narx}{\textit{Preprint} }
\newcommand{\arx}[1]{(\textit{Preprint} #1)}
\def\be{\begin{equation}}
\def\ee{\end{equation}}
\def\ba{\begin{eqnarray}}
\def\ea{\end{eqnarray}}
\def\bc{\begin{center}}
\def\ec{\end{center}}
\def\bs{\numparts}
\def\es{\endnumparts}
\def\a{\alpha}
\def\t{\tau}
\def\e{\epsilon}

\newcommand{\Eq}[1]{(\ref{#1})}

\begin{document}

\rightline{\small Class. Quantum Grav. \textbf{24} (2007) 829  \hfill gr-qc/0607059}

\title[Inflationary scalar spectrum in loop quantum cosmology]{Inflationary scalar spectrum in loop quantum\\ cosmology}
\author{Gianluca Calcagni and Marina Cort\^{e}s}
\address{Astronomy Centre, University of Sussex, Brighton BN1 9QH, United Kingdom}
\eads{\mailto{g.calcagni@sussex.ac.uk},\quad \mailto{m.v.cortes@sussex.ac.uk}}
\date{\today}

\begin{abstract}
In the context of loop quantum cosmology, we consider an inflationary era driven by a canonical scalar field and occurring in the semiclassical regime, where spacetime is a continuum but quantum gravitational effects are important. The spectral amplitude and index of scalar perturbations on an unperturbed de Sitter background are computed at lowest order in the slow-roll parameters. The scalar spectrum can be blue-tilted and far from scale invariance, and tuning of the quantization ambiguities is necessary for agreement with observations. The results are extended to a generalized quantization scheme including those proposed in the literature. Quantization of the matter field at sub-horizon scales can provide a consistency check of such schemes.
\end{abstract}

\pacs{98.80.Cq}

\maketitle


\section{Introduction}

One of the goals of modern theoretical physics is to find a consistent quantum description of gravity and the other interactions in Nature. This description should be able to reproduce the phenomenology of the standard particle and cosmological models and predict new, observable effects. So far, only two candidates seem to fulfil at least part of the necessary requirements of such an ambitious program. One is string theory: its nonperturbative structure is still under intense scrutiny but some of its contents were already exploited to motivate cosmological scenarios where the visible universe is an extended four-dimensional object (a brane) living in a higher-dimensional, continuous spacetime (see \cite{rub01,mar04,BVD,csa04} for some reviews).

An alternative, nonperturbative candidate is loop quantum gravity (LQG) \cite{rov97,thi02,smo04,rov04}, where the formulation of the quantum theory is carried out via the Hamiltonian formalism. The specialization to isotropic backgrounds, and in particular to a Friedmann--Robertson--Walker (FRW) metric, is referred to as loop quantum cosmology (LQC) (see \cite{boj06} for a review). The resulting spacetime is four-dimensional by construction.

The symmetry reduction of the theory is performed at the kinematical level, that is, before solving the Hamiltonian constraint \cite{boj1,boj2,boj3,boj4,ABL}. This procedure resembles standard minisuperspace quantizations, in the sense that the symplectic structure is reduced at the classical level \cite{ABL,APS}. However, the identification of the canonical variables and their quantization closely follow the formulation of loop quantum gravity. The states in the resulting Hilbert space, constructed in \cite{APS}, are in fact regarded as symmetric states of the full theory.

At very small scales, below a critical scale factor $a_{\rm i}$, spacetime is discretized and the Hamiltonian constraint equation can be written as an evolution equation with discrete time \cite{boj4,boj5}. Above another critical value $a_*$, larger than the Planck length $\ell_{\rm Pl}$, classical dynamics is recovered, while in the intermediate, semiclassical regime $a_{\rm i}<a\ll a_*$ spacetime can be treated as a continuum but nonperturbative quantum effects modify the standard cosmological evolution. The evolution equation can then be compared with the standard Wheeler--DeWitt equation \cite{boj4,boj5,boj7,boj9,boj10,DH2,SS1,SS2}.

Several authors have studied the semiclassical limit of LQC in the context of early-universe inflation \cite{boj9,boj10,DH2,hos03,TSM,boj11,boj12,DH1,hos04,NPV,lid04,BD,van05,sin05,SV,CLM,kag05,BK,MN}, or bouncing/cyclic scenarios \cite{APS,ST,LMNT,BMS,nun05,SVV}. Loop quantum gravity can leave its imprint on the large-scale structure of the universe and, in particular, the cosmic microwave background (CMB). However, the understanding of inflation in this framework is still partial and further study of its testable effects is required.

In this paper, we compute the main cosmological scalar observables from the perturbed background equations using standard techniques (e.g. \cite{lid97}), assuming that inflation takes place only in the semiclassical regime. The scalar amplitude and spectral index, as well as the index running, are found at lowest order in the slow-roll (SR) approximation, that is, on a de Sitter (dS) background. The scalar spectral index can be blue-tilted and far from the Harrison--Zel'dovich spectrum $n_{\rm s}=1$. We consider two examples of background homogeneous exact solutions and find that quasi scale invariance, as required by observations (see \cite{spe06} for WMAP experiment), is either lost or achieved at the price of a certain amount of tuning of the ambiguities.

While this work was under completion, other authors have considered the issue of scalar perturbations without metric backreaction \cite{MN}. We will compare their results with ours in due course.

The paper is organized as follows. In section~\ref{setup} we briefly recall basic concepts and construction of loop quantum cosmology. Several aspects and problems of LQC and the present treatment are listed in section~\ref{prob}. The inflationary scalar spectrum and index are found in section~\ref{obs} and calculated for a couple of examples in section~\ref{sol}. The analysis is extended to other quantization schemes in section~\ref{genqs}. Section~\ref{disc} is devoted to discussion of these results and future prospects.


\section{Setup}\label{setup}

The basic multiplication operator in homogeneous loop quantum gravity are point holonomies, which are $SU(2)$-elements associated with a single point on the space manifold and serve to describe quantum scalar fields. The further requirement of isotropy implies that quantum states (in the connection representation) are functions on a single copy of $SU(2)$. In particular, the holonomies can be written as $h_{\rm I}=\exp(\mu_0c\tau_{\rm I})$ in the spin-$1/2$ fundamental representation, where ${\rm I}=1,2,3$, $\tau_{\rm I}=-\rmi\sigma_\rmi/3 \in SU(2)$, $\sigma_\rmi$ are Pauli matrices, and $\mu_0=\sqrt{3}/4$ is a quantization ambiguity related to the length of the holonomy and fixed to a natural value \cite{ABL} (in many papers, $\mu_0=1$). The connection component $c=(\sqrt{K}-\gamma \dot a)/2$ is conjugate to the density weighted dreibein component $|p|=a^2$, where $K$ is the curvature constant ($K=0$ for a flat universe, $K=1$ for a closed one; the Hamiltonian formulation in the open case is treated separately \cite{AS,van06}), $\gamma\approx 0.2375$ is the Barbero--Immirzi parameter, and derivatives with respect to synchronous time are indicated as dots. The spatial volume is then ${\cal V}= |p|^{3/2}$ \cite{boj2}, while the symplectic structure is $\{c,p\}=\kappa^2\gamma/3$, where $\kappa^2\equiv 8\pi \ell_{\rm Pl}^2$ and curl brackets are Poisson brackets. 

The classical Hamiltonian constraint is
\be\label{clas}
{\cal H}=-\frac{12}{\kappa^2\gamma^2}\left[c(c-\sqrt{K})+(1+\gamma^2)\frac{K}{4}\right]\sqrt{|p|}+{\cal H}_{\rm mat}=0\,,
\ee
where ${\cal H}_{\rm mat}$ is the Hamiltonian of matter, which for simplicity we assume to be a real scalar field:
\be\label{clasm}
{\cal H}_{\rm mat}=\frac12 a^{-3} \pi_\phi^2+a^3 V(\phi)\,,
\ee
where $\pi_\phi=a^3\dot\phi$ is the momentum canonically conjugate to $\phi$ and $V$ is the field potential.

Volumes and densities are promoted to bounded operators with discrete spectra \cite{boj9,boj6,boj8}. The volume and dreibein operators $\hat{\cal V}$ and $\hat p$ have zero eigenvalue at the origin of the discretized `time' variable $t\in \mathbb{Z}$, and their inverses fail to be densely defined operators. However, the inverse-volume operator $\widehat{{\cal V}^{-1}}$ is regular (with zero eigenvalue) at $t=0$: this way the big bang singularity is healed by purely geometrical effects (see also \cite{BDH}).

The inverse volume can be written in a convenient way by using the classical identity $a^{-3}=[3(\kappa^2\gamma l)^{-1}\{c,|p|^l\}]^{3/(2-2 l)}$ \cite{boj8,boj11} and replacing the connection components with holonomies,
\be
a^{-3} = \left[\vphantom{\sum_\rmi}\frac{3}{\kappa^2\gamma\mu_0 lj(j+1)(2j+1)}\sum_\rmi {\rm tr}_j\left(\tau_\rmi h_\rmi\{h_\rmi^{-1},a^{2l}\}\right)\right]^{3/(2-2 l)},
\ee
which does not involve inverse, classically divergent powers of $a$ if $0<l<1$. 

The \emph{ambiguity parameter} $l$ determines the initial slope of the effective geometrical density. A natural choice, often used in literature, is $l=3/4$ \cite{boj12}. To preserve coordinate invariance when quantizing geometrical densities, $l$ must be discrete, $l_k=1-(2k)^{-1}$, $k\in \mathbb{N}$ \cite{thi97,boj12}. Hence one can select the bound $1/2\leq l <1$, which is also favoured phenomenologically \cite{boj12}.

The other ambiguity $j$ is a half-integer setting the value of the scale factor at which classical cosmology is recovered. It arises for arbitrary spin-$j$ representations (in which the above trace is taken) of the holonomy \cite{boj9,boj8}. The fundamental representation corresponds to $j=1/2$.

Promoting Poisson brackets to commutators of operators, one defines an inverse-volume operator. If the wavefunction of the Universe does not oscillate at small scales, difference operators can be approximated by differential operators which also appear in the standard Wheeler--DeWitt equation. The matter Hamiltonian, however, is not as in standard quantum cosmology, equation \Eq{clasm}, because of the promotion of the inverse volume to an operator $\hat d_{j,l}\equiv\widehat{a^{-3}}$ with eigenvalues $d_{j,l}\equiv D_l a^{-3}$, where
\ba
\fl D_l \equiv q^{3/2}\left\{\frac{3}{2l}\left[\frac{(q+1)^{l+2}-|q-1|^{l+2}}{l+2}-q\frac{(q+1)^{l+1}-{\rm sgn}(q-1)|q-1|^{l+1}}{l+1}\right]\right\}^{3/(2-2l)};\nonumber\\\label{dl}
\ea
here, $q\equiv (a/a_*)^2$, $a_*^2=a_\rmi^2j/3$ and $a_\rmi=\sqrt{\gamma\mu_0}\ell_{\rm Pl}$. Equation \Eq{dl} is valid for large $j$ and is a good approximation already at $j\gtrsim 10$. A large $j$ extends the period of superinflation \cite{boj12}, although this does not rule out values $1\lesssim j\lesssim 10^2$. As an upper bound, we note that no quantum gravitational effects have been detected so far in accelerators for energies below $E\sim 10^3$ GeV. Therefore the characteristic scale should be $(\sqrt{j}\ell_{\rm Pl})^{-1}< E$ and $j<10^{30}$.

$d_{j,l}$ has a maximum at $\approx a_*$. One recovers classical behaviour when $q\gg 1$ ($D_l\to 1$), while in the limit $q\ll 1$ (that is, in a regime where LQG effects are important) \cite{boj8,boj9}
\ba
D_l &\equiv& c^2_\nu\, a^{2(1-\nu)}\,,\label{Dlap}\\
    &\sim& \left(\frac{3}{l+1}\right)^{3/(2-2l)}\left(\frac{a}{a_*}\right)^{3(l-2)/(l-1)}.
\ea
In the first line we have defined the constant coefficient $c_\nu$ and
\ba
\nu \equiv 1-\frac12\frac{d\ln D_l}{d\ln a}=\frac{d\ln\a}{d\ln a}\,,
\ea
where
\be
\alpha \equiv a D_l^{-1/2}.
\ee
We will often encounter derivatives with respect to $\ln a$. This comes out naturally from the `intrinsic time formalism' of standard quantum gravity (as well as in the continuum limit of LQC, where the matter Hamiltonian is quantized differently \cite{boj4}), where $\ln a$ is the time variable of the Wheeler--DeWitt equation.

In the LQC regime 
\ba
\a  &\approx& \frac{a^\nu}{c_\nu},\label{aeff}\\
\nu &\approx& -\frac{4-l}{2(1-l)},\label{nueff}
\ea
while in the classical regime $\a\sim a$. Most of the equations below, unless stated otherwise (with a symbol $\approx$ as in equations \Eq{aeff} and \Eq{nueff}), are found for any smooth function $D_l$ and $\nu$.

It will be useful to note that the classical limit is achieved directly from equation \Eq{Dlap} when $\nu=1$ ($l=2$, $c_1=1$). The LQC regime is valid when $\nu<-2$, and the eigenvalues of the inverse volume operators are well defined even when $\nu\to -2$ ($l\to 0$); when $l=3/4$, $\nu\approx -13/2$. Hereafter $j>3$, so that $a_\rmi<a_*$ and the semiclassical regime is well defined.


\subsection{Equations of motion}

The semiclassical Hamiltonian is \cite{boj9}
\be\label{quh}
{\cal H}= -\frac{3}{\kappa^2}\,a(\dot a^2+K)+\frac12 d_{j,l} \pi_\phi^2+a(\nabla\phi)^2+a^3 V,
\ee
where now 
\be\label{mom}
\pi_\phi=d_{j,l}^{-1}\dot\phi\,,
\ee
and for later convenience we have included the gradient term. The inverse-volume operator and gravitational Hamiltonian are constructed through holonomies in the spin-$j$ and spin-$1/2$ representation, respectively. The constraint ${\cal H}=0$ is the Friedmann equation, while combining the Hamilton equations $\dot\phi=\{\phi,{\cal H}\}$ and $\dot\pi_\phi=\{\pi_\phi,{\cal H}\}$ one gets the dynamical equation for the scalar field:
\ba
&&H^2=\frac{\kappa^2}{3}\left(\frac{\dot\phi^2}{2D_l}+V\right)-\frac{K}{a^2}\,,\label{use1}\\
&& \ddot\phi-\frac{D_l}{a^2}\nabla^2\phi+(1+2\nu)H\dot\phi+D_l V_{,\phi}=0,\label{eom}
\ea
where $H\equiv \dot a/a$ is the Hubble parameter, and $V$ is differentiated with respect to $\phi$. At high energies the effective Klein--Gordon equation is that of a canonical field with potential $W=\int D_l V_{,\phi}\,\rmd\phi$ in a \emph{contracting} universe with $-(1+2\nu)>3$ dimensions. This simple remark eventually warns us that the resulting inflationary observables may be at odds with experiments, since the anti-friction term feeds the kinetic energy making it dominant over the potential energy \cite{TSM}.

The equations of motion can be recast in terms of an effective perfect fluid \cite{LMNT}:
\ba
\rho_{\rm eff} &\equiv& \frac{\dot\phi^2}{2D_l}+V\,,\label{eff1}\\
p_{\rm eff}    &\equiv& \frac{1+2\nu}{3}\frac{\dot\phi^2}{2D_l}-V\,,\label{eff2}
\ea
so that
\ba
&&H^2= \frac{\kappa^2}{3}\rho_{\rm eff}\,,\label{FEeff}\\
&&\dot\rho_{\rm eff}+3H(\rho_{\rm eff}+p_{\rm eff})=0\,.
\ea
In the limit $\nu\to 1$ these equations reduce to the Klein--Gordon energy density $\rho$ and pressure $p$.

From now on we move to the special case of a flat FRW background, $K=0$. 


\subsection{Quantization schemes}

As we shall see, ambiguities can in general be observed and fixed experimentally \cite{boj11,boj12,boj8}. While $l$ and $j$ appear at the kinematic level, other ambiguities arise when changing the quantization of the Friedmann equation. It is straightforward to extend the above equations of motion accordingly. 

Following the classification of \cite{boj12}, we denote with \textsc{Ham}$(n)$, $n\geq 0$, the scheme where arbitrary positive powers of $a^3 d_{j,l}$ are inserted into the Hamiltonian. Classically these insertions are equal to unity, but upon quantization they become dynamically nontrivial. In particular, in \textsc{Ham}$(n)$ both the matter effective energy density in the Friedmann equation and the definition of $\dot\phi$ in terms of the momentum $\pi_\phi$ are multiplied by $D_l^n$: $\rho_{\rm eff}(\pi_\phi,\phi)\to \rho^{(n)}_{\rm eff}(\pi^{(n)}_\phi,\phi) =D_l^n \rho_{\rm eff}(D_l^{-n}\pi_\phi,\phi)$. The above equations are in the \textsc{Ham}$(0)$ scheme. 

A second possibility is the \textsc{Fried} quantization \cite{hos03}, where the squared Hubble parameter is promoted to an operator and the Hamiltonian is no longer regarded as the primary object. When taking expectation values, the Friedmann equation reads $H^2\equiv \langle\widehat{H^2}\rangle\propto \langle\hat d_{j,l}\rangle\langle\hat{\cal H}_{\rm mat}\rangle$, and there appears an extra factor $D_l$ relative to the \textsc{Ham}$(0)$ Friedmann equation. The scalar momentum and field equation are unchanged.

A third case (which we will call \textsc{Pleb}) occurs when the classical gravitational Hamiltonian is derived from the self-dual Plebanski action instead from the Einstein--Hilbert one \cite{kag05}. Then ${\cal H}_{\rm Pleb}={\cal H}/\sqrt{p}$ classically (corresponding to a time reparametrization $\rmd t\to \sqrt{p}\,\rmd t$), which upon quantization gives extra factors of $D_l^{1/3}$.

Equations \Eq{use1} and \Eq{eom} can be generalized to all these schemes as
\ba
&&D_l^{n_1+1}H^2=\frac{\kappa^2}{3}\left(\frac{\dot\phi^2}{2}+D_l^{2n_2+1}V\right)\,,\label{usen}\\
&& \ddot\phi-\frac{D_l^{n_4+1}}{a^2}\nabla^2\phi+[(1+2\nu)+2n_3 (\nu-1)]H\dot\phi+D_l^{2n_2+1} V_{,\phi}=0,\label{eom2}
\ea
where $n_1$, $n_2$, and $n_3$ are shown in table \ref{tab1}. Here we have allowed also for a general correction to the gradient term, which was introduced in \cite{hos04} for $n_4=0$ (equation \Eq{quh}) on phenomenological grounds. Later on we shall see how to constrain it.
\begin{table}
\bc
\begin{tabular}{l|ccccc}
Scheme            & $n_1$ & $n_2$ & $n_3$ & $b_1$      & $b_2$        \\ \hline
\textsc{Ham}$(n)$ & $n$   & $n$   & $n$   & $(\nu-1)n$ & $2+\nu$      \\
\textsc{Fried}    & $-1$  & $0$   & $0$   & $\nu-1$    & $2+\nu$      \\
\textsc{Pleb}     & $1/3$ & $1/6$ & $1/3$ & $0$        & $(4\nu+5)/3$ \\
\end{tabular}\ec
\caption{\label{tab1}Ambiguities in different quantization schemes. The parameters $b_1$ and $b_2$ are defined in Sec.~\ref{genqs}.}
\end{table}

At first we shall consider in detail the case \textsc{Ham}$(0)$. It is interesting to check whether the results thus obtained are qualitatively robust across inequivalent quantization procedures, which is the case as argued in \cite{boj12}. The effect of the `secondary' ambiguities $n_\rmi$ is discussed in section \ref{genqs}.


\subsection{Superinflation and slow-roll parameters}

Using $\dot D_l/D_l=2H(1-\nu)$, one finds the Raychaudhuri equation:
\be\label{ray}
\dot H = -(2+\nu)\frac{\kappa^2}{6} \frac{\dot\phi^2}{D_l}.
\ee
Since $\nu+2<0$, the universe superaccelerates in the semi-classical phase ($\dot H>0$). 
The first parameters of the slow-roll tower are chosen as in the classical case:
\ba
\e   &\equiv& -\frac{\dot H}{H^2}\,,\\
\eta &\equiv& -\frac{\ddot\phi}{H\dot\phi}\,,\\
\xi^2&\equiv& \frac{1}{H^2}\left(\frac{\ddot\phi}{\dot\phi}\right)^. .
\ea
The first expression is purely geometrical, does not depend on the effective action, and has the advantage to define inflation precisely when $\e<1$. According to the second formula, the condition $|\eta|\ll \Or(|\nu|)$ is sufficient to neglect $\ddot\phi$ with respect to the modified friction term in the Klein--Gordon equation.

From equations \Eq{use1} and \Eq{ray}, the evolution equations of the SR parameters are
\ba
\dot\e &=& 2H\e\left[\e-\eta-(1-\nu)+\frac{\dot\nu}{2H(2+\nu)}\right]\label{dotex}\\
       &\approx& 2H\e[\e-\eta-(1-\nu)],\label{dothe}\\
\dot\eta &=& H(\e\eta-\xi^2).
\ea
The extra factor $(1-\nu)$ in equation \Eq{dothe} will contribute to modify the scalar spectrum.

Since we want to extract the cosmological observables at lowest order in the SR parameters, it is sufficient to assume a de Sitter background ($H=$ const). Note that in the traditional SR approximation all the observables are written at the same SR order, but for our purposes the dS computation will be enough, even if it does not allow such a systematic truncation (see section \ref{vali}). In particular, on can ignore all derivatives of $H$ and $V$ in the equations of motion and metric perturbations in the linearly perturbed equations.


\section{Caveats and open problems of semiclassical LQC}\label{prob}

Before proceeding, we underline several issues characterizing different formulations and approximations of loop quantum cosmology.


\subsection{Conformal rescaling}

In a flat FRW background, the scale factor can be rescaled by a conformal transformation which leaves the classical equations of motion invariant. If the spatial homogeneous slice is noncompact ($K=0$), all integrations must be performed over a fiducial cell, of size $\sim a$, within that slice. Then the fiducial cell also has the above `gauge' freedom (in a closed universe, one can fix the cell to have unit volume). On the other hand, the characteristic scale $a_*$ is fixed by the theory and the ratio $a/a_*$, as well as its functions and the statement that `the semiclassical regime is such that $a<a_*$', are not conformally invariant (\cite{APS}, appendix B.2). Then the theory and its observable implications would depend on the choice of $a$, i.e.~on specific initial conditions, even in the recent `$\bar\mu$ quantization' \cite{APS,sin06}. As we shall see, although the normalization of the scalar spectrum depends on $a_*$, the spectral index does not, and contains only the ambiguity $\nu$. However, both observables are derived in the semiclassical limit and the gauge dependence is, at best, implicit.

An escape route from the rescaling issue might be to start with a closed universe, for which there is no ambiguity in the scale factor. If the background is superaccelerating, the classical connection is $c=a(\sqrt{K/a^2}- \gamma H)\approx - \gamma aH$, since the first term rapidly vanishes. Then the universe loses memory about the initial curvature and one can use the equations in flat FRW. However, one may not be entitled to rely upon this picture too close to the singularity, when the scale factor, although expanding, is still very small; this might indeed happen in semiclassical LQC inflation. Also, there may be additional complications when perturbations are taken into account. This point should be addressed in the near future.


\subsection{Improved formulations of LQC}

The previous Hamiltonian constraint equation \Eq{quh}, in the spin-$1/2$ representation and in the flat case ($K=0$), was generalized through a WKB approximation directly at the level of the difference equation \cite{DH2,BD} and via a path integral quantization \cite{NPV,van05}. Corrections to the classical constraint are generated by the discrete structure of spacetime when smoothing the dynamical difference equations near the Planck scale to the continuum limit.
Upon quantization of the connection in terms of holonomy operators, the factor $c^2$ in equation \Eq{clas} is modified as $c^2\to \mu_0^{-2}\sin^2(\mu_0c)$ \cite{BD,SV} and the relation between $c$ and $\dot a$ becomes $\dot a=-(\gamma\mu_0)^{-1}\sin(\mu_0c)\cos(\mu_0c)$. The effective Friedmann equation is then 
\be\label{rrFE}
H^2 \propto \rho_{\rm eff}\left(1-\frac{\rho_{\rm eff}}{\rho_{\rm crit}}\right),
\ee
where $\rho_{\rm crit}$ is a critical density. This kind of evolution has applications for bouncing scenarios \cite{SVV}. 

At the time of completion of this paper, we assumed that inflation takes place at energies $\rho_{\rm eff}> \rho_{\rm crit}$, that is, when $\mu_0 c\ll 1$; then the standard Friedmann equation \Eq{FEeff} can be used as in the rest of the discussion. A problem might have been that recollapse can in principle occur during the inflationary regime \cite{BD}, since $\rho_{\rm crit}$ can be made arbitrarily small by increasing the scalar field momentum. However, the above `$\mu_0$ quantization' does not take into account the expansion of the fiducial cell in a FRW background. This was done in the novel $\bar\mu$ quantization described in \cite{APS,sin06}, where
\be
\mu_0 \to \bar\mu\propto \frac{\ell_{\rm Pl}}{a}\,.
\ee
The functional form of the inverse volume eigenvalues turns out to be modified; the Friedmann equation is still equation \Eq{rrFE} but with $\rho_{\rm crit}$ fixed and around 80\% of the Planck density. The equation for the scalar is the classical one after neglecting rapidly damped quantum corrections (\cite{APS}, appendix B). One would get a standard Friedmann equation when $\bar\mu c\ll 1$, that is, roughly when $H\ell_{\rm Pl}\ll 1$. This corresponds to the \emph{classical} regime, and all quantum corrections are neglected accordingly. Such a picture is qualitatively different from the semiclassical setup enunciated so far\footnote{Also, in the $\bar\mu$ version of LQC the quantum connection $c(\dot a,K)$ is not a linear function of $K$ \cite{singh}. The role of the curvature in expanding and bouncing scenarios is still to be assessed.}.

There is another, independent source of concern. The parameter $j$ is related to representation choices when quantizing inverse volumes in the matter Hamiltonian. A similar representation ambiguity arises also in the gravitational sector. Thus the classical term $\sqrt{|p|}$ becomes a function $s_{j,l}$ which is approximated by $a$ when $a>a_*$ ($l$ is fixed to $3/4$ \cite{van05,APS1} or $1/2$ \cite{APS}).

In order for the inverse volume corrections to have significant impact on the dynamics, one has to quantize the Hamiltonian using the representation of the gauge group different than fundamental ($j>1/2$). As a matter of fact, the semiclassical regime discussed above makes sense only in this `large $j$' representation; otherwise, there is no intermediate stage between the discrete quantum regime and the continuum classical limit.

For consistency, we should choose the same representation both for geometric objects in the gravitational term of the Hamiltonian and for those in the matter sector. However, in the higher $j$ case the Hamiltonian constraint is a difference equation of higher-than-second order. This may lead to an enlargement of the physical Hilbert space and, as a consequence, to the presence of solutions with incorrect large-volume limit \cite{van05}. Even if this were not the case, there is evidence (in 2+1 dimensions the proof is actually complete) that LQG has a well-defined continuum limit to quantum field theory only in the fundamental representation of the gauge group \cite{per05}. As said above, for small $j$ the semiclassical phase either is reduced or disappears altogether.

These issues are not completely settled but they would deeply modify the `old' quantization adopted here as working model, with corresponding change in the physical picture of the early universe\footnote{For recent insights on them, we refer the reader to \cite{bo609}. There, both the large-$j$ representation problem and the matter/gravity consistent representation problem are reinterpreted, and eventually relaxed, as natural features related to those of inhomogeneous backgrounds on a rigid lattice.}. Bearing them in mind, we shall try nonetheless to compute the power spectrum and the scalar spectral index. It would be interesting to compare our results in the absence of metric backreaction with those with full perturbations (initiated in \cite{BHKSS}; the same background equations are used). This would allow to verify what kind of approximations in the treatment of the cosmological problem preserve the physical features of the model.


\subsection{Large j and the number of e-foldings}

One can evaluate the maximum amount of superinflation during the semiclassical regime by 
requiring it to start at the Planck era ($a=a(t_\rmi)=a_\rmi$) and to end at $a(t_*)=a_*$. The number of $e$-foldings 
$N(t)\equiv\ln [a_{\rm end}/a(t)]$ is then
\be
N_{\rm max}=\ln (a_*/a_\rmi)=\frac12\ln(j/3)\,.
\ee
Therefore, $j\sim 10^{50}$ in order to get $N_{\rm max}\approx 57$, which is far beyond 
the allowed upper limit determined by particle physics. Then semiclassical superinflation 
alone is not sufficient to solve the flatness problem. One might invoke some mechanism 
lowering the minimum number of $e$-folds (an example in classical cosmology is thermal 
inflation \cite{LS1,LS2}) but $j$ would be still very large; for example, $N_{\rm max}
\approx 25$ requires $j\sim 10^{22}$, unrealistic in the quantum theory.

Assuming the universe accelerates even after the semiclassical-to-classical transition, 
would it be possible to see effects of an early semiclassical stage in the power spectrum?
The range of observable large-scale structures roughly spans the last ten $e$-foldings 
(e.g., \cite{LL}); these will pertain the classical period, and only predictions of standard 
inflation should be expected. The details of the model may still be tuned so that to get a semiclassical
modification of the scalar spectrum at low multipoles (very large scales); in particular, any deviation from scale 
invariance in the calculations below should be thought of as occurring in that region of 
the spectrum. 

Moreover, we note that, as argued in \cite{LPB}, the horizon and flatness 
problems are solved if the relative change in the comoving horizon $(aH)^{-1}$, not that of $a^{-1}$, is large 
enough. Namely, inflation is more precisely characterized by the number of $e$-foldings
\be
\bar N(t) \equiv \ln \frac{a_{\rm end}H_{\rm end}}{a(t) H(t)}\,.
\ee
In the standard case $H$ is almost constant and this definition coincides with the 
previous one. In the semiclassical case, it would enhance $N_{\rm max}$ by a term $\ln 
(H_*/H_\rmi)$, which is positive since $H$ increases in time. Hence, a smaller $j$ is 
required to get a certain number of $e$-folds.

In the case under scrutiny, we will compute the power spectrum under the assumption of a 
de Sitter background, later extending the result to a situation where the Hubble parameter 
can vary in time as $(H^{-1})^.\sim \Or(\nu)$ (see section \ref{vali}). Then superacceleration 
can in principle relax the high-$j$ problem.


\section{Perturbed equations and observables}\label{obs}

We perturb the scalar field around the homogeneous background $\phi_0(t)$, and substitute $\phi(\mathbf{x},t)=\phi_0(t)+\delta\phi(\mathbf{x},t)$ in equation \Eq{eom}. Going to momentum space, one has
\be
\delta\ddot{\phi}_\mathbf{k}+(1+2\nu)H\delta\dot\phi_\mathbf{k}+\left[\frac{D_lk^2}{a^2}+(D_l V_{,\phi})_{,\phi}\right]\,\delta\phi_\mathbf{k}=0,\label{peom}
\ee
where $H$, $D_l$ and $V$ depend only on time and a bold subscript indicates the $k$th Fourier mode. 

The dS approximation ($H\approx$ const) is self-consistent if
\be\label{ntsr}
|\e|,|\eta|\ll \Or(|\nu|)\,;
\ee
$|\eta|\ll \Or(|\nu|)$ is required to neglect the mass term in equation \Eq{peom}, proportional to
\ba
(D_l V_{,\phi})_{,\phi}&=&H^2\left[(1+2\nu)(\e+\eta)-\eta^2-\xi^2-\frac{2\dot\nu}{H}\right],\label{use2}
\ea
with respect to the mass given by equation \Eq{mass} below, $m_\nu^2\propto H^2 \nu(\nu+1)$. For $|\nu|\to \Or(1)$, these conditions are the usual SR constraints, which are too restrictive in this scenario when $|\nu|$ is large. 

The slow-roll condition $|\epsilon|\ll \Or(|\nu|)$ can be restated as $\dot\phi^2\ll D_l V$. Thus, in general, if the kinetic term dominates over the potential one has $|\epsilon|\sim \Or(|\nu|)$ (also, $D_l\ll 1$ in the semiclassical regime). Then one may wonder whether the anti-friction term easily causes the kinetic energy to overcome $D_l V$ and exit the slow-roll regime. Later on we shall discuss what happens if equation \Eq{ntsr} is relaxed to $|\e|,|\eta|\sim \Or(|\nu|)$, and argue that $\epsilon=$ constant is a sufficient consistency condition.


\subsection{Mukhanov equation}

It is convenient to change time coordinate and define a generalized `conformal time'\footnote{The name `conformal' is technically incorrect here, since the metric with this time coordinate is not equivalent to the Minkowski metric up to a conformal transformation: $ds^2=-\a^2d\t^2+a^2dx^\rmi dx_\rmi$. We shall keep it to distinguish from synchronous time $t$.}
\be\label{conf}
\rmd\t \equiv \frac{\rmd t}{\alpha},
\ee
which reduces to the standard conformal time in the classical limit. In the following, primes will denote derivatives with respect to $\t$. In analogy with the general relativistic case, one can define a Mukhanov variable $u_\mathbf{k}\equiv\a\,\delta\phi_\mathbf{k}$. Then equation \Eq{peom}, ignoring the term in $V$, becomes\footnote{One can also use the standard conformal time $\rmd\t=\rmd t/a$ and the above definition of $u_\mathbf{k}$ to get a Mukhanov equation $u_\mathbf{k}''+(D_lk^2-\a''/\a)u_\mathbf{k}=0$; its solution is (a linear combination of) $|u_\mathbf{k}|\propto \sqrt{-k\t}\,Z_{\pm \mu}(c_\nu \nu^{-1}H^{\nu-1}k\t^\nu)$ for any Bessel function $Z_\mu$ of order $\mu=\sqrt{(2\nu)^{-2}+1+\nu^{-1}}$ (\cite{GR}, formula 8.491.4). Again one obtains the power spectrum found below \cite{MN}. Also, the definition of horizon crossing $D_l k^2-\a''/\a\approx 0$ coincides with that given below, since $\a''/\a=\nu(\nu+1)(aH)^2$ in dS.}
\be\label{ueom}
u_\mathbf{k}''-f_\nu u_\mathbf{k}'+\left(k^2+m_\nu^2\right)u_\mathbf{k}=0,
\ee
where $k^2=|\mathbf{k}|^2$ and
\ba
f_\nu   &\equiv& \a H(\nu-1)\,,\\
m_\nu^2 &\equiv& -\frac{\a''}{\a}+\frac{\a'}{\a}f_\nu\\
&=& -\a H \nu\left[\a H(\nu+1) +\frac{H'}{H}+\frac{\nu'}{\nu}\right];\label{mass}
\ea
we have not yet assumed that $H$ and $\nu$ are constant and these expressions are exact. It is straightforward to see that in the classical limit ($D_l=$ const) $f_\nu=0$, $m_\nu^2=-a''/a$, and one recovers the standard Mukhanov equation.

Some useful formul\ae\ to get equation \Eq{ueom} are:
\bs\ba
&&\frac{\rmd}{\rmd t}=\frac1\alpha \frac{\rmd}{\rmd\t}\,,\qquad \frac{\rmd^2}{\rmd t^2}=\frac1{\alpha^2}\left(\frac{\rmd^2}{\rmd\t^2}-\frac{\alpha'}{\alpha}\frac{\rmd}{\rmd\t}\right),\\
&&\delta\phi'=\frac1\a \left(u'-\frac{\a'}\a u\right)\,,\\
&&\delta\phi''=\frac1\a \left[u''-2\frac{\a'}\a u'+\left(2\frac{{\a'}^2}{\a^2}-\frac{\a''}\a\right)u\right],\\
&& \frac{\a'}{\a}=\a H \nu\,,\quad \frac{\a''}{\a}=\frac{\a'}{\a}\left(2\frac{\a'}{\a}+\frac{H'}{H}+\frac{\nu'}{\nu}\right).
\ea\es

Equation \Eq{ueom} can be solved analytically only for very special functions $f_\nu(\t)$ and $m_\nu^2(\t)$ and at this point we limit the discussion to the LQC de Sitter regime with $\nu=$ const, where one can exactly integrate equation \Eq{conf} via equation \Eq{aeff} and get
\be\label{tau}
\t=-\frac1{\nu\a H}.
\ee
$\tau$ is an increasing function of synchronous time, running from 0 ($t=0$, vanishing scale factor) to $+\infty$ ($t\to +\infty$).
Then
\ba
f_\nu &\approx& \left(\frac1\nu-1\right)\frac1\t\,,\label{feff}\\
m_\nu^2 &\approx& -\left(\frac1\nu+1\right)\frac1{\t^2}\,,\label{meff}
\ea
in agreement with equation (30) of \cite{hos04}.

The solution is 
\be\label{exsol}
u_{\mathbf k} = B_{\nu,k}(-k\tau)^{1/(2\nu)} H_{1+1/(2\nu)}^{(1)}(-k\tau),
\ee
where $B_{\nu,k}$ is a constant and $H_{1+1/(2\nu)}^{(1)}$ is the Hankel function of the first kind of order $1+1/(2\nu)$ (see \cite{GR}, formula 8.491.3). We shall use the two asymptotic limits
\ba
H_\mu^{(1)}(x) &\sim& -\rmi\,\frac{\Gamma(\mu)}{\pi}\left(\frac{x}{2}\right)^{-\mu}\,,\qquad x\ll 1\,,\\
H_\mu^{(1)}(x) &\sim& \sqrt{\frac{2}{\pi x}}\,\, \rme^{\rmi\left(x-\frac\pi2\mu-\frac\pi4\right)}\,,\qquad x\gg 1\,.
\ea
where in the first line $\mu>0$ is assumed.

In the long wavelength limit $k/(\a H)\rightarrow 0$ ($-k\t\to 0$), when the mode with comoving wave number $k$ is outside the horizon, one has
\be \label{smallk}
|u_{\mathbf k}| \sim B_{\nu,k}\Gamma[1+1/(2\nu)]\frac{2^{1+1/(2\nu)}}{\pi}\left(-\frac{1}{k\t}\right)\,.
\ee
Here we have use the positivity of $3/4<1+1/(2\nu)<1$ to drop terms proportional to $(-k\t)^{1+1/(2\nu)}$.

At early times (short wavelengths, $k\t\to+\infty$), perturbation modes are still inside the Hubble horizon and asymptotically we have
\be \label{largek}
u_{\mathbf k} \sim B_{\nu,k}\sqrt{\frac{2}{\pi}}\,(-k\t)^{(1-\nu)/(2\nu)}\rme^{-\rmi k\t}\,,
\ee
up to a phase dependent on $\nu$.  When $\nu=1$ (large volumes), one recovers the Bunch--Davies vacuum solution (positive-frequency\footnote{This justifies a posteriori the choice made in equation \Eq{exsol} to set equal to zero the coefficient of the other independent solution $H_{1+1/(2\nu)}^{(2)}(-k\tau)$. This coefficient may be actually proportional to $(1-\nu)$, but we will keep equation \Eq{exsol} for simplicity.} plane waves in the Minkowski limit $H\to 0$) $u_{\mathbf k} \sim \rme^{-\rmi k\t}/\sqrt{2k}$ if, and only if, $B_{1,k}=\sqrt{\pi/(4k)}$. In general, the friction term in equation \Eq{ueom} does not vanish and the evolution of the particle perturbations is not driven by a wave equation. This is expected in the LQC framework, since small scales are those mainly sensitive to loop quantum effects\footnote{There are other theories where the adiabatic vacuum is not the most natural one, as in noncommutative and trans-Planckian models \cite{dan02}.}. Nevertheless, we show below that in the quantization scheme \textsc{Ham}$(1/2)$ the Bunch--Davies vacuum arises naturally.
 
We can determine $B_{\nu,k}$ by imposing standard commutation relations for the quantized scalar field. These read 
\bs\ba
&&\left[\hat\phi (\mathbf{x}_1,\,t),\,\hat\phi(\mathbf{x}_2,\,t)\right] = 0\,,\\
&&\left[\hat\pi_\phi (\mathbf{x}_1,\,t),\,\hat\pi_\phi(\mathbf{x}_2,\,t)\right]=0\,,\\
&&\left[\hat\phi (\mathbf{x}_1,\,t),\,\hat\pi_\phi(\mathbf{x}_2,\,t)\right] = \rmi\, \delta^{(3)}(\mathbf{x}_1-\mathbf{x}_2)\,,\label{comst}
\ea\es
where hats denote operators and the conjugate momentum $\pi_\phi$ is given by equation \Eq{mom}. In general, the composite operator $\hat\pi_\phi=\widehat{a^3\dot\phi}$ is not isomorphic to the composition of operators $\hat{d}_{j,l}^{-1}\widehat{\dot\phi}$. However, in the semiclassical regime the states of the Hilbert space on which the operators are defined can be decomposed into a gravitational and matter sector, $|\Psi\rangle=\sum_{\a,\beta}|{\rm grav}\rangle_\a\otimes |{\rm mat}\rangle_\beta$. Then geometrical and matter operators commute, and we can take the eigenvalue of the formers in equation \Eq{comst}. In other words, gravitational variables are treated as background while matter fields are quantized\footnote{In general, geometrical and matter operators do not act separately on physical states because solutions to the Hamilton constraint already incorporate correlations between the two sectors \cite{bojpc}. So operators on such states are in general complicated, entangled observables. The issue is delicate and will not be dealt with here.}.

In momentum space and time $\tau$, the nonvanishing commutator for the variable $u_\mathbf{k}$ is
\be\label{1}
\left[\hat u_{\mathbf{k}_1}(\t),\,\hat u_{\mathbf{k}_2}'(\t)\right] = \rmi D_l^{-1/2}\delta^{(3)}(\mathbf{k}_1-\mathbf{k}_2).
\ee
The Mukhanov variable is expanded in terms of creation and annihilation operators,
\be\label{co}
\hat u_\mathbf{k}=w_k a_\mathbf{k}+w_k^* a_\mathbf{k}^\dagger,
\ee
where $*$ indicates complex conjugate, $w_k$ is the classical solution previously denoted by $u_\mathbf{k}$, and
\bs\ba
&&[a_{\mathbf{k}_1},\,a_{\mathbf{k}_2}^\dagger] = \delta^{(3)}(\mathbf{k}_1-\mathbf{k}_2)\,,\label{3}\\
&&[a_{\mathbf{k}_1},\,a_{\mathbf{k}_2}] = 0 =[a_{\mathbf{k}_1}^\dagger,\,a_{\mathbf{k}_2}^\dagger]\,.
\ea\es
Plugging equation \Eq{co} into \Eq{1} and using equation \Eq{3}, one gets $w_k {w_k^*}'-w_k^*w_k'=\rmi D_l^{-1/2}$. The small-scale solution \Eq{largek} satisfies this constraint only if
\be\label{norm}
B_{\nu,k}^2=\frac\pi2\frac{(-k\t)^{1-1/\nu}}{2k D_l^{1/2}}.
\ee
This is indeed a constant in $\t$, as one can check from equations \Eq{Dlap}, \Eq{aeff} and \Eq{tau}\footnote{This choice differs from that in \cite{hos04}; see equation (45) in that paper. There, the different normalization together with the ansatz $k_*=aH$ for horizon crossing lead to ${\cal P}_\phi\propto H^2$.}.

The spectrum of the scalar field fluctuations is
\be\label{Pphi}
{\cal P}_{\phi}\equiv \frac{k^3}{2\pi^2}\left\langle|\delta\phi_{\mathbf k}|^2\right\rangle\Big|_{k=k_*},
\ee
where angular brackets denote the vacuum expectation value of the perturbation on the state defined as $a_\mathbf{k}|0\rangle=0$ $\forall\,\mathbf{k}$. Conventionally, perturbative modes cross the Hubble horizon when $k^2=k_*^2\approx -m_\nu^2 =\nu(\nu+1)(\a H)^2$ from equations \Eq{tau} and \Eq{meff}. The positive coefficient in front of $\a H$ can be included in the overall normalization of the spectrum, and we can set the wave number at crossing as
\be\label{ansat}
k_*\equiv\a H.
\ee
The solution plugged into equation \Eq{Pphi} is that at long wavelengths, equations \Eq{smallk} and \Eq{norm}, although it is evaluated at horizon crossing. Then, up to a numerical factor (equal to 1 when $\nu=1$),
\be\label{spec}
{\cal P}_{\phi}=\frac{1}{\sqrt{D_l}}\left(\frac{H}{2\pi}\right)^2,
\ee
where $D_l$ is given by equation \Eq{Dlap} with constant $\nu$. Clearly, this quantity is not constant in dS due to the presence of $D_l$.

The ansatz \Eq{ansat} will be crucial to get the correct expression of the spectral index. Horizon crossing occurs at scales much smaller than in the standard scenario, $k_*\sim D_l^{-1/2}(aH)\gg \tilde k\equiv aH$. We note also that, while in the standard case comoving modes appear to exit the horizon, ($\rmd|\tilde k|^{-1}/\rmd t<0$), in this scenario they enter it, since the comoving effective horizon $|k_*|^{-1}$ expands, $\rmd|k_*|^{-1}/\rmd\tau>0$. This raises an important interpretational issue \cite{MN}, the limit $k\tau\to 0$ being not a natural consequence of the inflationary evolution. However, this does not affect the power-law solution below, which we shall take as a viable LQC example. This solution is not de Sitter ($H\neq$ const) but we will argue in section \ref{vali} that it can be taken as a background for the dS perturbations, as happens in standard inflation. Bearing in mind that the horizon-crossing problem still remains, for the moment we let it aside and push further the analytic inspection of the model. \emph{A posteriori} this allows us to find one puzzling result which might be relevant in future developments of quantum gravity: namely, quantization schemes may be constrained by requiring consistent (pseudo) plane-wave perturbations at small scales.


\subsection{Comoving curvature perturbation}

The spectrum of scalar perturbations is the two-point correlation function of the curvature perturbation on comoving flat slices, defined as in classical gravity
\be\label{R}
{\cal R}=\frac{H}{\dot{\phi}}\,\delta\phi\,,
\ee
to linear order. We are going to show this result by aid of the covariant approach, first introduced in \cite{EB,BDE,BED} and recently refined in \cite{LV1,LV2}. The starting point is a inhomogeneous spacetime with metric $g_{\mu\nu}$ (signature $({-}{+}{+}{+})$, Greek indices from 0 to 3) and filled with a perfect fluid characterized by the energy-momentum tensor $T_{\mu\nu}=(\rho+p)u_\mu u_\nu+p g_{\mu\nu}$, where $u_\mu=\rmd x_\mu/\rmd t$ is the fluid four-velocity ($u_\mu u^\mu=-1$, $u_\mu \dot u^\mu=0$) and $t$ is proper time along flow lines. For a Klein--Gordon scalar field with $\rho=-\nabla_\sigma\phi\nabla^\sigma\phi/2+V$ and $p=-\nabla_\sigma\phi\nabla^\sigma\phi/2-V$, the energy-momentum tensor is $T_{\mu\nu}=\nabla_\mu\phi\nabla_\nu\phi+p g_{\mu\nu}$, which compared with the above expression gives 
\be\label{ust}
u_\mu=-\frac{\nabla_\mu\phi}{\sqrt{-\nabla_\sigma\phi\nabla^\sigma\phi}}==-\frac{\nabla_\mu\phi}{\dot\phi}\,,
\ee
which is orthogonal to $\phi=$ const hypersurfaces. Here, $\dot{}=\partial_t=u^\sigma\nabla_\sigma$. One introduces
\ba
N &\equiv& \frac13\int \rmd t \nabla_\sigma u^\sigma\\
&=& \int \rmd t\, H,
\ea
where $\nabla_\sigma$ is the covariant derivative and in the last line we have specialized to a FRW background; on it, $N(t)$ is just the number of $e$-foldings. However, here $N(\mathbf{x},t)$ and $H(\mathbf{x},t)$ are time and space dependent. The variable of interest is the covector ${\cal R}_\mu\equiv -D_\mu N$, where $D_\mu\equiv (\delta_\mu^\nu+u_\mu u^\nu)\nabla_\nu=\nabla_\mu+u_\mu\partial_t$ is the spatial gradient. Via equation \Eq{ust}, one has
\be\label{Rex}
R_\mu=-\partial_\mu N+\frac{H}{\dot\phi}\partial_\mu\phi\,,
\ee
where all variables are functions of both time and space. This quantity is the nonlinear generalization of the perhaps more familiar linear comoving curvature perturbation ${\cal R}$, and satisfies a conservation equation at all scales thanks to the conservation of the energy momentum tensor, $\nabla^\mu T_{\mu\nu}=0$, without use of the Einstein equations. At linear order, ${\cal R}_\rmi\approx \partial_\rmi {\cal R}$, where the index i runs over spatial coordinates; in the flat gauge $\delta N=0$, ${\cal R}$ is given by equation \Eq{R}, with $H$ and $\dot\phi$ now coinciding with the background homogeneous quantities.

In loop quantum cosmology, a covariantly conserved effective stress-energy tensor for a perfect fluid is
\be\label{Teff}
T_{\mu\nu}^{\rm eff}=(\rho_{\rm eff}+p_{\rm eff})\,u_\mu u_\nu+p_{\rm eff} g_{\mu\nu}\,,
\ee
with $\rho_{\rm eff}$ and $p_{\rm eff}$ given by equations \Eq{eff1} and \Eq{eff2} in FRW for the scalar field. Then one can start from a classical Einstein--Hilbert action and get the Friedmann equation \Eq{FEeff} as the 00 component of the Einstein equations $G_{\mu\nu}=\kappa^2T_{\mu\nu}^{\rm eff}$. Equation \Eq{Teff} corresponds to
\be\label{Tefflqc}
T_{\mu\nu}^{\rm eff}=\frac{2+\nu}{3D_l} \nabla_\mu\phi\nabla_\nu\phi+p_{\rm eff} g_{\mu\nu}\,.
\ee
Equation \Eq{Tefflqc} is covariant if $D_l$ is regarded as a function of the volume ${\cal V}$. The simplest way to implement the covariant formalism in LQC is to promote all the above quantities (including $D_l$) to time- and space-dependent variables in a locally defined `separate' universe \cite{WMLL}. It is not obvious and likely nontrivial to show that this construction is compatible with loop quantum gravity; we shall assume it as working hypothesis, leaving its check for the future. In this case, we get again equations \Eq{ust}, \Eq{Rex} and \Eq{R}.

A remark is in order. Within the same framework, one can introduce the uniform density perturbation (at linear order) $\zeta_{\rm eff}$ defined on uniform density slices where $\delta\rho_{\rm eff}=0=\delta p_{\rm eff}$. This perturbation will be different from the uniform density perturbation $\zeta$ on slices where $\delta\rho=0=\delta p$, and while the latter is equivalent to ${\cal R}$ (up to a sign and at large scales), the former is not. The reason is that, for a canonical Klein--Gordon field, $\delta\rho-\delta p =2V_{,\phi}\delta\phi$. The left-hand side vanishes on uniform density slices, implying that these are also comoving ($\delta\phi=0$). On the other hand, $\delta\rho_{\rm eff}-\delta p_{\rm eff}=2V_{,\phi}\delta\phi+2(1-\nu)\dot\phi(\delta\dot\phi+C\dot\phi)/3$, where the term in $C$ comes from the perturbed metric in the kinetic term $g_{\mu\nu}\partial^\mu\phi\partial^\nu\phi$. If $\nu\neq 1$, in general the right-hand side is nonvanishing on comoving slices. Then, ${\cal R}\approx-\zeta$ and $\zeta_{\rm eff}$ are inequivalent candidates for the observable spectrum. We skip further subtleties inherent to this choice (including the question about conservation of these quantities) and opt for the comoving curvature perturbation. The author of \cite{hos04} chose the density contrast $\delta\rho/\rho$.


\subsection{Scalar spectrum and index}

The scalar spectrum is 
\bs\ba
A_{\rm s}^2 &\equiv& \frac{2k^3}{25\pi^2} \left\langle |{\cal R}_{\mathbf k}|^2\right\rangle\Big|_{k=k_*}\\
&=& \left(\frac25\frac{H}{\dot\phi}\right)^2{\cal P}_{\phi}\,.\label{Scsp}
\ea\es
We have chosen a normalization which allows a transparent comparison with the standard result; this can be achieved by compensating the normalization of the solution for the Mukhanov equation (fixed by commutation relations) with a suitable, approximate definition of horizon crossing. Changing the latter would lead to a $\nu$-dependent factor in front of the amplitude \Eq{Scsp}, which can in principle be tested by CMB observations. This would enlarge the parameter space and loosen individual constraints on other variables.

The final result reads
\bs\ba
A_{\rm s}^2 &=&  \frac{1}{\sqrt{D_l}}\left(\frac{1}{5\pi}\frac{H^2}{\dot\phi}\right)^2\\
      &=&  \frac{(2+\nu)\kappa^2}{6(5\pi)^2}\frac{H^2}{D_l^{3/2}\e}.\label{as}
\ea\es
The spectral index is defined as
\be
n_{\rm s}-1 \equiv \frac{\rmd \ln A_{\rm s}^2}{\rmd \ln k}\Big|_{k=k_*}\,,
\ee
where at horizon crossing $\rmd\ln k =(\dot\a/\a)\rmd t=\nu H\rmd t$ on dS. From equations \Eq{as} and \Eq{dothe},
\be
n_{\rm s}-1 = \frac1\nu [n_{\rm s}^{\rm(cl)}-1-(1-\nu)],
\ee
where $n_{\rm s}^{\rm(cl)}-1=2\eta-4\e$ is the standard scalar index. Heuristically, since $\nu<0$ the scalar index can be \emph{blue tilted} if the classical one is red tilted. 

This finding agrees qualitatively with that achieved by the `direct method' of \cite{hos04}, although comparison of the expressions for the scalar index (equation (43) of \cite{hos04}) is not straightforward. A blue tilt corresponds to a loss of power at large scales. This can be understood by the introduction of a cutoff scale, of order $\ell_{\rm Pl}$, below which no modes are produced from vacuum (this is nothing but the `curvature cutoff' provided by quantization of inverse powers of the scale factor). Such a cutoff is ideally present also in the standard picture, but here it is rather close to the energies involved ($a\gtrsim\ell_{\rm Pl}$). A conclusion similar to that of \cite{hos04} was reached in another minisuperspace cosmological model \cite{HW}, where the canonical variables and quantum Hilbert space and operators are not that of LQG (see \cite{HuW} for details). Note that also this model has an ambiguity analogous to $l$, which was fixed in \cite{HW,HuW}.

However, we will show that the parameter space and freedom in the ambiguities allow for both red-tilted and blue-tilted spectral indices, while the scalar index found in \cite{hos04} is always blue tilted regardless the value of $\nu$. Finally, the running of the spectral index is
\be
\a_{\rm s}\equiv \frac{\rmd n_{\rm s}}{\rmd\ln k}=\frac2{\nu^2}[5\e\eta-4\e^2-\xi^2+4\e(1-\nu)].
\ee
The last term is $\Or(\e)$ and may dominate over the others.


\section{Background solutions and the SR approximation}\label{sol}


\subsection{Power-law solution}

\begin{figure}[t]
\bc
{\psfrag{nu}{$\nu$}
\psfrag{lam}{$\lambda$}
\psfrag{n1}[hb][][.8][0]{$n_{\rm s}-1$}
\includegraphics[width=8cm]{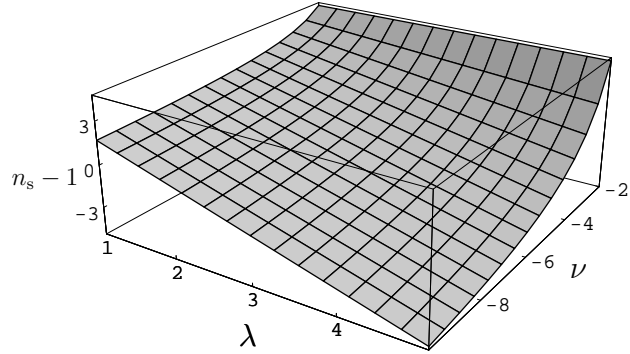}
\caption{\label{fig1} The scalar spectral index \Eq{sppl} as a function of the parameter $\lambda>1$ and the ambiguity $\nu$.}}\ec
\end{figure}

\begin{figure}
\bc{\psfrag{nu}{$\nu$}
\psfrag{lam}{$\lambda$}
\includegraphics[width=8cm]{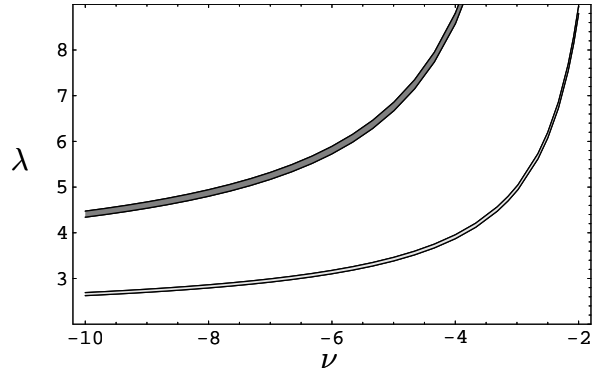}
\caption{\label{fig2} Region in the $\lambda$-$\nu$ plane predicting a spectral index $0.9<n_{\rm s}<1.0$ for the \textsc{Ham}$(0)$ (upper, dark strip) and \textsc{Pleb} (lower, light strip) quantization scheme (see section \ref{genmu}).}}\ec
\end{figure}

A power-law solution of the background equations was found in \cite{lid04} when $D_l$ is given by equation \Eq{Dlap}:
\bs\ba
a(t)    &=& t^{1/[(2+\nu)\lambda]},\\
\phi(t) &=& \phi_\nu\, t^{(1-\nu)/[(2+\nu)\lambda]},\\
V(\phi) &=& V_\nu\, \phi^{-2(2+\nu)\lambda/(1-\nu)},
\ea\es
where (after time reversal) the constant $\lambda>1$ in order to have a classically stable expanding solution and $\phi_\nu$ and $V_\nu$ are appropriate normalization constants. The quadratic potential $V\propto \phi^2$ considered in \cite{TSM} is stable in the LQC regime ($\lambda>1$ for $\nu<-2$). The SR parameters read
\bs\ba
\e   &=& (2+\nu)\,\lambda\,,\\
\eta &=& (2+\nu)\,\lambda -(1-\nu)\,,\\
\xi^2&=& (2+\nu)\lambda[(2+\nu)\lambda-(1-\nu)].
\ea\es
The observable spectral index is constant ($\a_{\rm s}=0$):
\be\label{sppl}
n_{\rm s}-1 = 3\left(1-\frac1\nu\right)-2\left(1+\frac2\nu\right)\lambda,
\ee
and is shown in figure \ref{fig1} as a function of $\nu$ and $\lambda$. Most of the parameter space generates a strongly scale-dependent spectrum. To highlight a region compatible with observations (figure \ref{fig2}, upper strip) we require $-0.1<n_{\rm s}-1<0$, which has the best-fit $n_{\rm s}\approx 0.95$ of WMAP3 \cite{spe06} as central value. The scalar spectrum can be blue tilted in the semiclassical LQC regime but one can tune the parameters to get a red-tilted, almost scale-invariant spectrum. In the figures we have considered the conservative interval $0<l<1$ instead of $1/2\leq l<1$ ($\nu\leq -7/2$).

Note that the $k_*^{-1}$ actually decreases in time as $\tau$, and modes exit the effective horizon.


\subsection{Exponential solution}

Another exact solution is of exponential type and can be found either by direct calculation or via the mapping found in \cite{lid04} from the standard solution:
\bs\ba
a(t) &=& \exp\left(\frac{3p}{2+\nu}t^s\right)\,,\\
\dot\phi(t) &=& \dot\phi_\nu\, t^{s/2-1}\exp\left[\frac{3(1-\nu)}{2+\nu}pt^s\right]\,,\\
V(t) &=& V_\nu\, t^{2(s-1)}(1+A_\nu t^{-s}),
\ea
where $p$ and $s\neq 1$ are constants and
\ba
V_\nu &=& \frac{3}{\kappa^2}\left(\frac{3ps}{2+\nu}\right)^2,\\
\dot\phi_\nu^2 &=&\frac{3}{\kappa^2}\frac{6 ps(1-s)c_\nu^2}{(2+\nu)^2}\,,\\
A_\nu &=& -\frac{\dot\phi_\nu^2}{2c_\nu^2V_\nu}.
\ea\es
Reality of the scalar field requires $ps(1-s)>0$; setting $\nu=1$ we recover the classical solution \cite{bar90,BS}. 

The SR parameters read
\bs\ba
\e    &=& \frac{(2+\nu)(1-s)}{3ps}\,t^{-s},\\
\eta  &=& (\nu-1)+\frac{(2+\nu)(2-s)}{6ps}\,t^{-s},\\
\xi^2 &=& \left(\frac{2+\nu}{3ps}\right)^2\left[\left(1-\frac{s}{2}\right)t^{-s}+\frac{3ps(s-1)(1-\nu)}{2+\nu}\right]t^{-s}.\nonumber\\
\ea\es
For positive $s$, these quantities decay in time and the spectral index is asymptotically $n_{\rm s}-1\sim 3(1-\nu^{-1})$; hence $3<n_{\rm s}-1<4.5$, which is strongly blue tilted and excluded by observations.


\subsection{Validity of the de Sitter solution and Lidsey mapping}\label{vali}

The solution equation \Eq{exsol} was derived under the assumption that both $\e$ and $\eta$ are small enough, $\ll \Or(|\nu|)$ (massless field approximation). The last examples, however, show that $|\e|$ and $|\eta|$ can be $\Or(|\nu|)$. There may be doubts about the self-consistency of the spectral index thus found. 

In standard inflation, $\epsilon\ll 1$ is assumed for the calculation of the dS spectrum into which, nonetheless, one inserts a solution (power law) which has $\epsilon$ constant of arbitrary magnitude (if the constant is small, one has scale invariance). The observables for power-law inflation are not at odds with the dS calculation, since the latter agrees with the lowest-order perturbative calculation carried out under the assumption $\e,\eta=$ const.

As in the standard case, an exact solution of the LQC Mukhanov equation at next-to-leading SR order can be achieved when $\e,\eta=$ const, no matter their magnitude. By virtue of equation \Eq{use2}, corrected with an extra geometric term in $\e$ from the backreaction of the metric (which is presently out of the discussion), the friction and mass terms \Eq{feff} and \Eq{meff} would be replaced by $[2\chi_2(\e,\eta,\nu)-1]/\tau$ and $[\chi_2^2(\e,\eta,\nu)-\chi_3^2(\e,\eta,\nu)]/\tau^2$ respectively, where $\chi_2$ and $\chi_3$ are constants built with the slow-roll parameters. The solution would be a linear combination of $(-k\tau)^{\chi_2} Z_{\chi_3}(-k\tau)$, where $Z_{\chi_3}$ is a Bessel function, and will have different asymptotic behaviours relative to equation \Eq{exsol}. The spectrum at horizon crossing, to lowest SR order, is unchanged anyway (see the following section, equation \Eq{spec2} with $\chi_0=1$). This might be also required by robustness of black-hole results, as hinted in the concluding section.

The key point is that the adopted `slow-roll' approximation is actually a `not-too-fast-roll' one, defined in equation \Eq{ntsr}. Roughly speaking, the requirement for the dS approximation to hold is much looser in LQC than in classical general relativity, in the sense that it is consistent even when the field velocity is not very small. This further 
legitimates the choice of solutions which are not de Sitter, even if the calculations assume $H=$ constant.

The SR parameters can be derived also by using the duality of \cite{lid04}, which relates solutions $\{a(t),H(t),\phi(t)\}$ of the LQC equations to solutions of standard general relativity with scale factor $b(t)$, Hubble parameter $\beta=\dot b/b$ and a Klein--Gordon scalar $\psi(t)$, where $b = a^{3/(2+\nu)}$, $\beta = (2+\nu)H/3$, and $\dot\psi = -(2+\nu) \dot\phi/(3\sqrt{D_l})$. The SR parameters calculated for the standard solution $\{b(t),\beta(t),\psi(t)\}$ are
\ba
\tilde\e &\equiv& -\frac{\dot\beta}{\beta^2}=\frac3{2+\nu}\,\e\,,\\
\tilde\eta &\equiv& -\frac{\ddot\psi}{\beta\dot\psi}=\frac3{2+\nu}[\eta+(1-\nu)].
\ea
Hence, a dS solution in LQC corresponds to a standard solution with $\tilde\e=0$ and $\tilde\eta=\nu-1$; this partly explains deviation from scale invariance in the examples.

Also the scalar spectrum \Eq{spec} can be achieved via the above mapping. One starts from the dual system of a Klein--Gordon field in standard general relativity. The two-point correlation function of the fluctuation $\delta\psi$ is evaluated at the standard horizon crossing, $\tilde k=aH$, and is given by the Gibbons--Hawking temperature, ${\cal P}_\psi=\beta^2/(2\pi)^2\propto H^2$. One argues that 
\be\label{specmap}
{\cal P}_\phi\propto k_*^3\langle|\delta\phi|^2\rangle\propto (D_l^{-1/2}\tilde k)^3 D_l\langle|\delta\psi|^2\rangle
\propto \frac{{\cal P}_\psi}{\sqrt{D_l}}\,,
\ee
in agreement with equation \Eq{spec}. 


\section{Cosmology in a generalized quantization scheme}\label{genqs}

Having discussed the implications of LQC for the scalar spectrum in the simple \textsc{Ham}$(0)$ quantization, we can easily extend the analysis to the other schemes. The background and perturbed dS dynamics can be written starting from the general equations \Eq{usen} and \Eq{eom2}. In order to preserve the standard continuity and Friedmann equations, equations \Eq{eff1} and \Eq{eff2} become
\ba
\rho_{\rm eff} &\equiv& \frac{\dot\phi^2}{2D_l^{n_1+1}}+D_l^{2n_2-n_1}V\,,\label{eff3}\\
p_{\rm eff}    &\equiv& \left[\frac{2}{3}(b_1+b_2)-1\right]\frac{\dot\phi^2}{2D_l^{n_1+1}}-\left(1-\frac{2}{3}b_1\right)D_l^{2n_2-n_1}V\,,\label{eff4}
\ea
where
\ba
b_1 &\equiv& (1-\nu)(n_1-2n_2),\\
b_2 &\equiv& (\nu+2)+2(n_2-n_3)(1-\nu).
\ea
In the usual scheme \textsc{Ham}$(0)$, $b_1=0$ and $b_2=\nu+2$ (see table \ref{tab1}). For \textsc{Ham}$(n\neq 0)$, $b_1,b_2<0$, while in \textsc{Pleb} $b_1=0$ and $b_2<-1$. The Raychaudhuri equation is
\be\label{ray2}
b_1H^2+\dot H = -b_2\frac{\kappa^2}{6} \frac{\dot\phi^2}{D_l^{n_1+1}}.
\ee
The evolution equation for $\epsilon$ now reads
\ba
\dot\e &=&\dot b_1+2H(\e-b_1)\left[\e-\eta-(1+n_1)(1-\nu)+\frac{\dot b_2}{2Hb_2}\right]\\
       &\approx& 2H(\e-b_1)[\e-\eta-(1+n_1)(1-\nu)],
\ea
in agreement with equation \Eq{dotex}.


\subsection{Mukhanov equation and scalar field spectrum}\label{genmu}

In analogy with what has been done so far, one might define a `conformal' time and canonical variable $u\equiv \tilde\a\delta\phi\equiv a D_l^{-(1+n_1)/2}\delta\phi$. However, the physical result does not depend on the choice of time, and we can keep the old definition of $\tau$ without loss of generality.

The gradient term in the Mukhanov-type equation has an extra factor $D_l^{n_4}$, which can be written as a function of $\t$ by using the formula $D_l\approx c_\nu^{2/\nu}(-\nu H\t)^{2(1-1/\nu)}$, from equations \Eq{Dlap} and \Eq{tau}. The friction and mass terms are
\bs\ba
f_\nu &=&\a H (\nu-1)(1-2n_3)\label{fnu}\\
&\approx& \left(\frac1\nu-1\right)\frac{1-2n_3}\t,\\
m_\nu^2 
&=& -\a H \nu\left\{\a H[2\nu+(1-\nu)(1-2n_3)]+\frac{H'}{H}+\frac{\nu'}{\nu}\right\}\\
&\approx& -\left[2+\left(\frac1\nu-1\right)(1-2n_3)\right]\frac{1}{\t^2}.
\ea\es
The Mukhanov equation and its solution are
\be
u_{\mathbf k}''-\frac{2\chi_2-1}{\t}\,u_{\mathbf k}'+\left[(\chi_0\chi_1\t^{\chi_0-1})^2+\frac{\chi_2^2-\chi_0^2\chi_3^2}{\t^2}\right]u_{\mathbf k}=0\,,
\ee
\be
u_{\mathbf k} = B_{\nu,n_\rmi,k}(-\chi_1\tau)^{\chi_2} H_{\chi_3}^{(1)}(-\chi_1\tau^{\chi_0}),
\ee
where
\bs\ba
\chi_0 &\equiv& \left(1-\frac1\nu\right)n_4+1\to 1\,,\\
\chi_1 &\equiv& \left[c_\nu^{1/\nu}(-\nu H)^{1-1/\nu}\right]^{n_4}\frac{k}{\chi_0}\to k\,,\\
\chi_2 &\equiv& \frac{1-2n_3}{2\nu}+n_3\to \frac1{2\nu}\,,\\
\chi_3 &\equiv& \frac{\chi_2+1}{\chi_0}\to 1+\frac{1}{2\nu}\,,
\ea\es
and the limits are taken as $n_\rmi\to 0$ for all $\rmi$. Horizon crossing is defined as 
\be
\chi_{1*}\approx (\a H)^{\chi_0}\quad\Rightarrow\quad k_*\approx\a^{\chi_0} H\,.
\ee
At large and small scales, respectively,
\ba
|u_{\mathbf k}| &\sim& B_{\nu,n_\rmi,k}\frac{\Gamma(\chi_3)}{\pi}2^{\chi_3}(-\chi_1)^{\chi_2-\chi_3}\tau^{\chi_2-\chi_0\chi_3},\\
u_{\mathbf k} &\sim& B_{\nu,n_\rmi,k} \sqrt{\frac{2}{\pi}}\,(-\chi_1)^{\chi_2-1/2}\tau^{\chi_2-\chi_0/2}\rme^{-\rmi\chi_1\tau^{\chi_0}}.
\ea
The momentum conjugate to $\phi$ is $\pi_\phi=a^3D^{-(1+n_1)}\dot\phi$ and the commutation relation \Eq{1} becomes
\be\label{1bis}
\left[\hat u_{\mathbf{k}_1}(\t),\,\hat u_{\mathbf{k}_2}'(\t)\right] = \rmi D_l^{n_1-1/2}\delta^{(3)}(\mathbf{k}_1-\mathbf{k}_2).
\ee
Imposing equation \Eq{1bis}, the integration constant of the small-scale solution is
\be
B_{\nu,n_\rmi,k}^2=\frac\pi2\frac{(-\chi_1\t^{\chi_0})^{1-2\chi_2}D_l^{n_1-1/2}}{2\chi_1\chi_0},
\ee
which is independent of $\tau$ if the consistency condition $\chi_0(1-2\chi_2)+(2n_1-1)(1-1/\nu)=0$ is satisfied:
\be
2(n_1-n_3)+\left(1-\frac1\nu\right)(1-2n_3)n_4=0.
\ee
This constrains the unknown parameter $n_4$ in almost any quantization scheme.
\begin{itemize}
\item \textsc{Ham}$(n)$: It must be $n_4=0$ for any $n\neq1/2$, while for $n=1/2$ any value of $n_4$ is allowed.
\item \textsc{Fried}: Only $2/3<n_4=2\nu/(\nu-1)<1$ is allowed. Since $n_4$ should be independent of the ambiguity $\nu$, it is unlikely that the \textsc{Fried} scheme would lead to a consistent quantization at small scales.
\item \textsc{Pleb}: Only $n_4=0$ is allowed.
\end{itemize}
These results can be interpreted in different ways:
\begin{itemize}
\item The commutation relation \Eq{1bis} is indeed a selection rule on the possible quantization schemes.
\item The dS approximation is not reliable and an expansion in constant SR parameters for the \emph{full} perturbed equations (i.e.~including metric backreaction), or a different approach altogether, is required.
\item Any allowed quantization scheme in the symmetry reduced theory is legitimate and one has to give up either the commutation relation \Eq{comst} or the oscillator expansion \Eq{co}.
\end{itemize}
Due to the evident limitations of our method, we do not try to solve this issue here. It will be important to address it when further developments are available.

In the \textsc{Ham}$(1/2)$ scheme with $n_4=0$, $2\chi_2-1=0$ and the short wavelength solution is a plane wave, $u_\mathbf{k}\sim \rme^{-\rmi k\tau}$, thus recovering the Bunch--Davies vacuum. This would be also true in the general case $2n_3-1=n_4$, after a time redefinition $\tau^{\chi_0}\to \tau$, but such condition is not compatible with the above selection rules.

The scalar spectrum ${\cal P}_\phi$ is
\ba
{\cal P}_\phi &=& D_l^{n_1-1/2}\frac{(k/\alpha)^2}{(2\pi)^2} \frac{\Gamma^2(\chi_3)}{\pi\chi_0}2^{2\chi_3-1}\nonumber\\
&&\qquad\times\,k\chi_1^{-(\chi_2+\chi_3)}\t^{\chi_0(1-2\chi_2-\chi_3)+\chi_2}\Big|_{k=k_*}\\
&\propto& D_l^{n_1-1/2}\left(\frac{H}{2\pi}\right)^2(\a^{\chi_2+2}H^{\chi_2-1})^{\chi_0-1},\label{spec2}
\ea
in agreement with the previous result ${\cal P}_\phi(k)\propto D_l^{-1/2} (k/\a)^2(k\t)^{-1/\nu}$ and equation \Eq{spec}. The spectrum in \textsc{Ham}$(1/2)$ with $n_4=0$ is precisely the standard one.

The comoving curvature perturbation was discussed in the \textsc{Ham}$(0)$ case. The only modification is in the form of the effective stress-energy tensor \Eq{Teff}, which is now constituted by the energy density \Eq{eff3} and pressure \Eq{eff4} with $\dot\phi^2\to -\nabla_\sigma\phi\nabla^\sigma\phi$.


\subsection{Background solutions, generalized spectrum and mapping}

The exponential solution has an unacceptably blue-tilted spectrum and we shall ignore it. The LQC power-law solution is
\bs\ba
a          &=& t^{1/\e}\,,\\
\phi &\propto& t^{(1-\nu)(n_1+1)/\e}\,,\\
V    &\propto& t^{2(1-\nu)(n_1-2n_2)/\e-2}\,,\\
\eta &=& \e-(1-\nu)(n_1+1)\,,\\
\xi^2 &=& \e[\e-(1-\nu)(n_1+1)].
\ea\es
When $b_1=0$ as in \textsc{Ham}$(0)$ and \textsc{Pleb}, one can extend the duality of \cite{lid04} as follows:
\ba
\beta &=& \frac{b_2}{3} H\,,\\
\dot\psi &=& -\frac{b_2}3 \frac{\dot\phi}{D_l^{(n_1+1)/2}}\,,\label{newdu}
\ea
and an appropriate redefinition of the potential $V(\phi)\to W(\psi)$ which is not important for the discussion. Then the Raychaudhuri equation \Eq{ray2} can be written as the standard one:
\be\label{ray3}
\dot\beta=-\frac{\kappa^2}{2}\dot\psi^2,
\ee
with a Friedmann equation $\beta^2\propto \rho_\psi=\dot\psi^2/2+ W(\psi)$. This determines $\e = b_2\lambda$ in terms of the parameter $\lambda$ from the standard power-law solution $a\sim t^{1/(3\lambda)}$. 

It is instructive to see how different quantization schemes affect the scalar spectrum. When $b_1=0$, then $\chi_0=1$ and the spectral index reads
\be
n_{\rm s}-1 = \frac1{\nu}[-2b_2\lambda-(2n_1+3)(1-\nu)]\,.
\ee
The \textsc{Ham}$(0)$ and \textsc{Pleb} cases are shown in figure \ref{fig2}. The two schemes have similar qualitative features.

When $b_1\neq 0$, the relation between the Hubble parameters $H$ and $\beta$ is not algebraic and, in order to preserve equations \Eq{newdu} and \Eq{ray3}, one has to define
\be\label{defi}
\dot\beta\equiv\frac{b_2}3(\dot H+b_1 H^2)=\frac{b_2}3\dot H\left(1-\frac{b_1}{\e}\right)\,.
\ee
An alternative might have been to modify equation \Eq{newdu} so that to simplify the mapping $H \mapsto \beta$, but that is the only definition allowing to interpret the spectrum \Eq{spec2} in terms of the dual, standard spectrum for a Klein--Gordon field. By the same procedure of equation \Eq{specmap}, one has that ${\cal P}_\phi \propto D_l^{n_1-1/2} \beta^2$. For all backgrounds with constant $\e$, $\beta$ is proportional to $H$.

Since the mapping is between homogeneous backgrounds, the next step is to get rid of those terms in equation \Eq{spec2} which bear the explicit imprint of the gradient term. In the limit $\chi_0\to 1$, the spectrum agrees with that found via the above line of reasoning.

The full calculation of the scalar spectrum, however, takes into account at least part of the inhomogeneities due to quantum fluctuations, and its expression is more complicated than what would be naively expected from the sole mapping argument. Hence the relation, as formulated here, between standard and LQC scenarios involves only background solutions and not observables.

In the power-law case, $\beta=(3\lambda t)^{-1}$, $H=(\e t)^{-1}$ and equation \Eq{defi} is quadratic in $\e$. It yields
\be
\e=\frac{b_2\lambda}{2}\left(1\pm \sqrt{1-\frac{4b_1}{b_2\lambda}}\right),
\ee
which is real if $\lambda\geq 4b_1/b_2$. From table \ref{tab1}, in the two cases of interest \textsc{Ham}$(n\neq 0)$ and \textsc{Fried}, the solution is well-defined when $\lambda >4n$ and $\lambda >4$, respectively, thus restricting the stability condition $\lambda>1$.

At present we do not know how to fix $\chi_0$ in the \textsc{Ham}(1/2) quantization scheme, but obviously it would further change the allowed parameter space. However, the tuning on the parameters would not be relaxed but shifted to other regions.


\section{Conclusions}\label{disc}

To summarize, we have analysed the spectrum of cosmological observables generated during a period of superacceleration in the semiclassical regime of loop quantum cosmology, formulated in the `old' quantization schemes \textsc{Ham}$(n)$, \textsc{Fried} and \textsc{Pleb}. At the level of the background LQC equations, the following assumptions were done and discussed: flat metric, classical (or small) connection, large $j$ representation of geometrical objects in the matter Hamiltonian, different representation of geometrical objects in the gravitational part. By perturbing the background equations without taking into account the backreaction of the metric (i.e., on a pure de Sitter spacetime), the scalar spectrum of power-law inflation results scale-dependent, unless one tunes the parameters of the theory (including ambiguities). The short-wavelength behaviour of the perturbed solutions can constrain the ambiguities within each quantization scheme.

This does not mean that loop quantum cosmology is ruled out by observations. The first thing one should check is whether the semi-analytic results and the dS approximation agree with a full numerical solution of the Mukhanov equation \Eq{ueom} with $D_l$ given by equation \Eq{dl}. Another key point is that we have assumed that inflation occurs only in the semiclassical regime. In \cite{TSM,boj12} the LQC semiclassical regime (with $l=3/4$) was regarded as a selecting mechanism for initial conditions before classical inflation, with an intermediate, classical, SR-violating phase in between. This transient period may be then responsible for loss of power at large scales, and an indirect signature of loop quantum gravity if the field reaches maximum displacement from the potential minimum at least 60 $e$-folds before the end of inflation (here we have seen that low-momenta suppression of the spectrum can be also a direct LQC effect). It would be interesting to extend the analyses of \cite{TSM,boj12} to other inflaton potentials with general ambiguity parameter $l$ and $j$ relaxed to even small values.

On the other hand, if inflation takes place when the discrete structure of spacetime cannot yet be approximated by a continuum, one might be able to put nontrivial constraints on the theory by considering the full LQC regime and its difference equations. However, one would require explicit expressions for quantum observables which have not yet been determined. The task seems difficult but worth being explored.

Another trend of research might try to extend the matter Hamiltonian to a more general content. In \cite{BK}, for instance, the scalar field is assumed to be nonminimally coupled to gravity. Scalars not satisfying a Klein--Gordon equation at the classical level can produce qualitatively different, and possibly more viable, stages of inflation. We note, however, that certain alternative choices presently in fashion, like the Dirac--Born--Infeld tachyon, are motivated mainly by string theory, and their application to LQC \cite{sen06} should be regarded as purely phenomenological.

One of the most remarkable results of LQG is the proof of the Bekenstein--Hawking area law by counting the nonperturbative quantum states of a nonextremal black hole, a computation which fixes the Barbero--Immirzi parameter \cite{ABCK,kra98,KM,ABK,DL,mei04,AEV}. In general, it was shown that the area law is valid for all isolated horizons, including the cosmological ones, in the limit of large area ($H^{-1}>10^3 \ell_{\rm Pl}$ or so). Then the entropy of a dS horizon with radius $H^{-1}$ is $S \propto H^{-2}$. By interpreting the Friedmann equation $H^2\propto\rho_{\rm eff}$ as the (integrated) first law of dS thermodynamics $-\rmd E=T\rmd S$, where $\rmd E\propto H^{-3} \rmd\rho_{\rm eff}$ is an infinitesimal energy exchange through the horizon \cite{BR,FrK,CK}, the dS temperature is indeed the Gibbons--Hawking temperature $T_{\rm H}=H/(2\pi)$ \cite{GH1}. The spectrum of a Klein--Gordon field is ${\cal P}_\phi=T_{\rm H}^2$. 

It will be important to verify the thermodynamical interpretation of the Friedmann equation also for small volumes as in the semiclassical regime. In that case, logarithmic corrections to the entropy $S\sim H^{-2}+\log H+\dots$ should be taken into account. In fact, during semiclassical inflation one cannot use the large-volume expression $S \propto H^{-2}$, because $H^{-1}< a\ll a_*\sim \ell_{\rm Pl}$ for $j\lesssim 10^3$ (for larger $j$ there may be an allowed window but at this stage the interpretation of ${\cal P}_\phi$ as a modified Hawking temperature is not clear). Considerations on the canonical partition function of a quantum system in equilibrium suggest that the above holographic principle might hold in general, also in this regime, because the bulk Hamiltonian (and not the boundary one) is a constraint for the spectrum of physical states \cite{maj06}.

Although the de Sitter approximation (that is, $\phi$ as a test field with no backreaction from the metric) is generally a good one in standard inflation, it is not obvious that it is as well justified in LQC. Actually, the symmetry reduction of the full theory is performed under the very special assumption of homogeneity and isotropy. But even small inhomogeneous perturbations can have drastic effects \cite{boj2}, and quantization of an anisotropic Hamiltonian may lead to a rather different set of equations and spectra. Important open issues to be considered in much greater detail are the choice of the (vacuum?) state on which to evaluate the two-point correlation function of the scalar perturbation, the choice of gauge (is the flat gauge $\delta N=0$ consistent?) and gauge-invariant observables, the adiabaticity of perturbations, and the behaviour of the tensor spectrum, yet to be determined altogether. The impact of geometrical backreaction should be assessed and what has been shown in this paper be checked by a more rigorous calculation in order to be considered reliable. If this were the case, the standard technique we adopted would prove to be a useful, relatively simple tool of analysis for loop quantum cosmology.


\ack

GC was supported by Marie Curie Intra-European Fellowship MEIF-CT-2006-024523, and MC by FCT (Portugal). We thank A R Liddle, D J Mulryne, P Majumdar and P Singh for useful discussions, and M Bojowald for careful reading of the original manuscript and stimulating insights.



\section*{References}

\begin{thebibliography}{50}

\bibitem{rub01} Rubakov V A, 2001 \textit{Phys. Usp.} \textbf{44} 871 \arx{hep-ph/0104152}
\bibitem{mar04} Maartens R, 2004 \textit{Living Rev. Relativity} \textbf{7} 7 \arx{gr-qc/0312059}
\bibitem{BVD}   Brax P, van de Bruck C and Davis A-C, 2004 \textit{Rept. Prog. Phys.} \textbf{67} 2183 \arx{hep-th/0404011}
\bibitem{csa04} Cs\'{a}ki C, 2004 \narx{hep-ph/0404096}

\bibitem{rov97} Rovelli C, 1998 \textit{Living Rev.\ Rel.} {\bf 1} 1 \arx{gr-qc/9710008}
\bibitem{thi02} Thiemann T, 2003 \textit{Lect.\ Notes Phys.} {\bf 631} 412003 \arx{gr-qc/0210094}
\bibitem{smo04} Smolin L, 2004 \narx{hep-th/0408048}
\bibitem{rov04} Rovelli C, 2004 \textit{Quantum Gravity} (Cambridge: Cambridge University Press)

\bibitem{boj06} Bojowald M, 2005 \textit{Living Rev.\ Rel.} {\bf 8} 11 \arx{gr-qc/0601085}

\bibitem{boj1}  Bojowald M, 2000 \textit{Class.\ Quantum Grav.} {\bf 17} 1489 \arx{gr-qc/9910103}
\bibitem{boj2}  Bojowald M, 2000 \textit{Class.\ Quantum Grav.} {\bf 17} 1509 \arx{gr-qc/9910104}
\bibitem{boj3}  Bojowald M, 2001 \textit{Class.\ Quantum Grav.} {\bf 18} 1055 \arx{gr-qc/0008052}
\bibitem{boj4}  Bojowald M, 2001 \textit{Class.\ Quantum Grav.} {\bf 18} 1071 \arx{gr-qc/0008053}
\bibitem{ABL}   Ashtekar A, Bojowald M and Lewandowski J, 2003 \textit{Adv.\ Theor.\ Math.\ Phys.} {\bf 7} 233 \arx{gr-qc/0304074}
\bibitem{APS}   Ashtekar A, Pawlowski T and Singh P, 2006 \textit{Phys.\ Rev.} D {\bf 74} 084003 \arx{gr-qc/0607039}
\bibitem{boj5}  Bojowald M, 2001 \textit{Phys.\ Rev.\ Lett.} {\bf 87} 121301 \arx{gr-qc/0104072}
\bibitem{boj7}  Bojowald M, 2002 \textit{Class.\ Quantum Grav.} {\bf 19} 2717 \arx{gr-qc/0202077}
\bibitem{boj9}  Bojowald M, 2002 \textit{Phys.\ Rev.\ Lett.} {\bf 89} 261301 \arx{gr-qc/0206054}
\bibitem{boj10} Bojowald M and Vandersloot K, 2003 \textit{Phys.\ Rev.} D {\bf 67} 124023 \arx{gr-qc/0303072}
\bibitem{DH2}   Date G and Hossain G M, 2004 \textit{Class.\ Quantum Grav.} {\bf 21} 4941 \arx{gr-qc/0407073}
\bibitem{SS1}   Shojai F and Shojai A, 2006 \textit{Europhys.\ Lett.} {\bf 75} 702 \arx{gr-qc/0607033}
\bibitem{SS2}   Shojai F and Shojai A, 2006 \textit{Gen.\ Rel.\ Grav.} {\bf 38} 1387 \arx{gr-qc/0607034}
\bibitem{hos03} Hossain G M, 2004 \textit{Class.\ Quantum Grav.} {\bf 21} 179 \arx{gr-qc/0308014}
\bibitem{TSM}   Tsujikawa S, Singh P and Maartens R, 2004 \textit{Class.\ Quantum Grav.} {\bf 21} 5767 \arx{astro-ph/0311015}
\bibitem{boj11} Bojowald M, 2004 \textit{Pramana} {\bf 63} 765 \arx{gr-qc/0402053}
\bibitem{boj12} Bojowald M, Lidsey J E, Mulryne D J, Singh P and Tavakol R, 2004 \textit{Phys.\ Rev.} D {\bf 70} 043530 \arx{gr-qc/0403106}
\bibitem{DH1}   Date G and Hossain G M, 2005 \textit{Phys.\ Rev.\ Lett.} {\bf 94} 011301 \arx{gr-qc/0407069}
\bibitem{hos04} Hossain G M, 2005 \textit{Class.\ Quantum Grav.} {\bf 22} 2511 \arx{gr-qc/0411012}
\bibitem{NPV}   Noui K, Perez A and Vandersloot K, 2005 \textit{Phys.\ Rev.} D \textbf{71} 044025 \arx{gr-qc/0411039}
\bibitem{lid04} Lidsey J E, 2004 \textit{J. Cosmol. Astropart. Phys.} JCAP12(2004)007 \arx{gr-qc/0411124}
\bibitem{BD}    Banerjee K and Date G, 2005 \textit{Class.\ Quantum Grav.} {\bf 22} 2017 \arx{gr-qc/0501102}
\bibitem{van05} Vandersloot K, 2005 \textit{Phys.\ Rev.} D {\bf 71} 103506 \arx{gr-qc/0502082}
\bibitem{sin05} Singh P, 2005 \textit{Class.\ Quantum Grav.} {\bf 22} 4203 \arx{gr-qc/0502086}
\bibitem{SV}    Singh P and Vandersloot K, 2005 \textit{Phys.\ Rev.} D {\bf 72} 084004 \arx{gr-qc/0507029}
\bibitem{CLM}   Copeland E J, Lidsey J E and Mizuno S, 2006 \textit{Phys.\ Rev.} D {\bf 73} 043503 \arx{gr-qc/0510022}
\bibitem{kag05} Kagan M, 2005 \textit{Phys.\ Rev.} D {\bf 72} 104004 \arx{gr-qc/0511007}
\bibitem{BK}    Bojowald M and Kagan M, 2006 \textit{Phys.\ Rev.} D {\bf 74} 044033 \arx{gr-qc/0606082}
\bibitem{MN}    Mulryne D J and Nunes N J, 2006 \textit{Phys.\ Rev.} D {\bf 74} 083507 \arx{astro-ph/0607037}

\bibitem{ST}    Singh P and Toporensky A, 2004 \textit{Phys. Rev.} D \textbf{69} 104008 \arx{gr-qc/0312110}
\bibitem{LMNT}  Lidsey J E, Mulryne D J, Nunes N J and Tavakol R, 2004 \textit{Phys.\ Rev.} D {\bf 70} 063521 \arx{gr-qc/0406042}
\bibitem{BMS}   Bojowald M, Maartens R and Singh P, 2004 \textit{Phys. Rev.} D \textbf{70} 083517 \arx{hep-th/0407115}
\bibitem{nun05} Nunes N J, 2005 \textit{Phys.\ Rev.} D {\bf 72} 103510 \arx{astro-ph/0507683}
\bibitem{SVV}   Singh P, Vandersloot K and Vereshchagin G V, 2006 \textit{Phys.\ Rev.} D {\bf 74} 043510 \arx{gr-qc/0606032}

\bibitem{lid97} Lidsey J E, Liddle A R, Kolb E W, Copeland E J, Barreiro T and Abney M, 1997 \textit{Rev. Mod. Phys.} \textbf{69} 373 \arx{astro-ph/9508078}
\bibitem{spe06} Spergel D N {\it et al}, 2006 \narx{astro-ph/0603449}
\bibitem{AS}    Ashtekar A and Samuel J, 1991 \textit{Class. Quantum Grav.} \textbf{8} 2191
\bibitem{van06} Vandersloot K, 2006 \narx{gr-qc/0612070}

\bibitem{boj6}  Bojowald M, 2001 \textit{Phys.\ Rev.} D {\bf 64} 084018 \arx{gr-qc/0105067}
\bibitem{boj8}  Bojowald M, 2002 \textit{Class.\ Quantum Grav.} {\bf 19} 5113 \arx{gr-qc/0206053}
\bibitem{BDH}   Bojowald M, Date G and Hossain G M, 2004 \textit{Class.\ Quantum Grav.} {\bf 21} 3541 \arx{gr-qc/0404039}

\bibitem{thi97} Thiemann T, 1998 \textit{Class.\ Quantum Grav.} {\bf 15} 1281 \arx{gr-qc/9705019}
\bibitem{sin06} Singh P, 2006 \textit{Phys.\ Rev.} D {\bf 73} 063508 \arx{gr-qc/0603043}
\bibitem{singh} Singh P, 2006 private communication
\bibitem{APS1}  Ashtekar A, Pawlowski T and Singh P, 2006 \textit{Phys.\ Rev.} D \textbf{74} 084003 \arx{gr-qc/0604013}
\bibitem{per05} Perez A, 2006 \textit{Phys.\ Rev.} D \textbf{73} 044007 \arx{gr-qc/0509118}
\bibitem{bo609} Bojowald M, 2006 \textit{Gen. Rel. Grav.} {\bf 38} 1771 \arx{gr-qc/0609034}
\bibitem{BHKSS} Bojowald M, Hern\'{a}ndez H H, Kagan M, Singh M and Skirzewski A, 2006 \textit{Phys.\ Rev.} D {\bf 74}  123512 \arx{gr-qc/0609057}
\bibitem{LS1}   Lyth D H and Stewart E D, 1995 \textit{Phys.\ Rev.\ Lett.} \textbf{75} 201 \arx{hep-th/9502417}
\bibitem{LS2}   Lyth D H and Stewart E D, 1996 \textit{Phys.\ Rev.} D {\bf 53} 1784 \arx{hep-ph/9510204}
\bibitem{LL}    Liddle A R and Lyth D H, 2000 \textit{Cosmological inflation and large-scale structure} (Cambridge: Cambridge University Press)
\bibitem{LPB}   Liddle A R, Parsons P and Barrow J D, 1994 \textit{Phys.\ Rev.} D {\bf 50} 7222 \arx{astro-ph/9408015}
\bibitem{GR}    Gradshteyn I S and Ryzhik I M, 2000 \textit{Tables of Integrals, Series, Products} (San Diego: Academic Press)
\bibitem{dan02} Danielsson U H, 2002 \textit{Phys.\ Rev.} D \textbf{66} 023511 \arx{hep-th/0203198}
\bibitem{bojpc} Bojowald M, 2006 private communication

\bibitem{EB}    Ellis G F R and Bruni M, 1989 \textit{Phys.\ Rev.} D \textbf{40} 1804
\bibitem{BDE}   Bruni M, Dunsby P K S and Ellis G F R, 1992 \textit{Astrophys.\ J.} \textbf{395} 34
\bibitem{BED}   Bruni M, Ellis G F R and Dunsby P K S, 1992 \textit{Class.\ Quantum Grav.} \textbf{9} 921
\bibitem{LV1}   Langlois D and Vernizzi F, 2005 \textit{Phys.\ Rev.\ Lett.} \textbf{95} 091303 \arx{astro-ph/0503416}
\bibitem{LV2}   Langlois D and Vernizzi F, 2005 \textit{Phys.\ Rev.} D \textbf{72} 103501 \arx{astro-ph/0509078}
\bibitem{WMLL}  Wands D, Malik K A, Lyth D H and Liddle A R, 2000 \textit{Phys.\ Rev.} D \textbf{62} 043527 \arx{astro-ph/0003278}

\bibitem{HW}    Hofmann S and Winkler O, 2004 \narx{astro-ph/0411124}
\bibitem{HuW}   Husain V and Winkler O, 2004 \textit{Phys.\ Rev.} D {\bf 69} 084016 \arx{gr-qc/0312094}
\bibitem{bar90} Barrow J D, 1990 \textit{Phys. Lett.} B \textbf{235} 40
\bibitem{BS}    Barrow J D and Saich P, 1990 \textit{Phys. Lett.} B \textbf{249} 406
\bibitem{sen06} Sen A A, 2006 \textit{Phys.\ Rev.} D {\bf 74} 043501 \arx{gr-qc/0604050}

\bibitem{ABCK}  Ashtekar A, Baez J, Corichi A and Krasnov K, 1998 \textit{Phys.\ Rev.\ Lett.} {\bf 80} 904 \arx{gr-qc/9710007}
\bibitem{kra98} Krasnov K, 1998 \textit{Class.\ Quantum Grav.} {\bf 15} L47 \arx{gr-qc/9803074}
\bibitem{KM}    Kaul R K and Majumdar P, 2000 \textit{Phys.\ Rev.\ Lett.} {\bf 84} 5255 \arx{gr-qc/0002040}
\bibitem{ABK}   Ashtekar A, Baez J C and Krasnov K, 2000 \textit{Adv.\ Theor.\ Math.\ Phys.} {\bf 4} 1 \arx{gr-qc/0005126}
\bibitem{DL}    Domaga{\l}a M and Lewandowski J, 2004 \textit{Class.\ Quantum Grav.} {\bf 21} 5233 \arx{gr-qc/0407051}
\bibitem{mei04} Meissner K A, 2004 \textit{Class.\ Quantum Grav.} {\bf 21} 5245 \arx{gr-qc/0407052}
\bibitem{AEV}   Ashtekar A, Engle J and Van Den Broeck C, 2005 \textit{Class.\ Quantum Grav.} {\bf 22} L27 \arx{gr-qc/0412003}

\bibitem{BR}    Bak D and Rey S-J, 2000 \textit{Class.\ Quantum Grav.} \textbf{17} L83 \arx{hep-th/9902173}
\bibitem{FrK}   Frolov A and Kofman L, 2003 \textit{J. Cosmol. Astropart. Phys.} JCAP05(2003)009 \arx{hep-th/0212327}
\bibitem{CK}    Cai R-G and Kim S P, 2005 \textit{J. High Energy Phys.} JHEP02(2005)050 \arx{hep-th/0501055}
\bibitem{GH1}   Gibbons G W and Hawking S W, 1977 \textit{Phys.\ Rev.} D \textbf{15} 2738
\bibitem{maj06} Majumdar P, 2006 \narx{gr-qc/0604026}

\end{thebibliography}
\end{document}